\documentclass[12pt, a4paper]{article}
\pdfoutput=1
\usepackage{graphicx}
\usepackage{amssymb}
\usepackage{amsmath}              
\usepackage{bm}
\usepackage{color}
\usepackage{cases}% added by RN -14.11.19-
\usepackage{cite}
\usepackage{subfigure}

\newcommand{\Slash}[1]{{\ooalign{\hfil/\hfil\crcr$#1$}}}

\usepackage{hyperref}

\begin{document}

\begin{titlepage}

\begin{flushright}

FTPI-MINN-15/03 \\
IPMU15-0012

\end{flushright}

\vskip 2cm
\begin{center}

{\large
{\bf 
Effective Theories for Dark Matter Nucleon Scattering
}
}

\vskip 2cm

Junji Hisano$^{a,b,c}$,
Ryo Nagai$^b$,
and
Natsumi Nagata$^{c,d}$

\vskip 0.5cm

{\it $^a$Kobayashi-Maskawa Institute for the Origin of Particles and the
Universe, \\ Nagoya University, Nagoya 464-8602, Japan}\\[2pt]
{\it $^b$Department of Physics,
Nagoya University, Nagoya 464-8602, Japan}\\[2pt]
{\it $^{c}$Kavli Institute for the Physics and Mathematics of the Universe (WPI),\\
  The University of Tokyo Institutes for Advanced Study, The
University of Tokyo, Kashiwa, Chiba 277-8583, Japan}\\[2pt]
{\it $^{d}$William I. Fine Theoretical Physics Institute, School of
 Physics and Astronomy, University of Minnesota, Minneapolis, MN 55455,
 USA}

\date{\today}

\vskip 1.5cm

\begin{abstract} 

 We reformulate the calculation of the dark matter-nucleon scattering
 cross sections based on the method of effective field theories. We
 assume that the scatterings are induced by the exchange of colored
 mediators, and construct the effective theories by integrating out the
 colored particles. All of the leading order matching conditions as well
 as the renormalization group equations are presented. We consider
 {a Majorana fermion, and real scalar and vector bosons for the dark matter} and show the results for each
 case. The treatment for the twist-2 operators is discussed in detail,
 and it is shown that the scale of evaluating their nucleon matrix
 elements does not have to be the hadronic scale. The effects of
 the QCD corrections are evaluated on the assumption that the masses of
 the colored mediators are much heavier than the electroweak scale. Our
 formulation is systematic and model-independent, and thus suitable to
 be implemented in numerical packages, such as {\tt micrOMEGAs} and {\tt
 DarkSUSY}.

\end{abstract}

\end{center}
\end{titlepage}

%%%%%%%%%%%%%%%%%%%%%%%%%%%%%%%%%%%
\section{Introduction}
%%%%%%%%%%%%%%%%%%%%%%%%%%%%%%%%%%%%

Weakly Interacting Massive Particles (WIMPs) have been widely regarded
as the most attractive candidate for dark matter (DM) in the
Universe. They are weakly coupled to the Standard Model (SM)
particles so that they are thermalized in the early Universe. Their relic
abundance is determined by their annihilation cross sections, and it
turns out that WIMPs may have a correct value of the cross sections to
give the observed DM density $\Omega_{\text{DM}}h^2 =0.1196\pm
0.0031$~\cite{Ade:2013zuv}. Moreover, such particles are often predicted
in new physics beyond the SM. For instance, the lightest neutralino in
the minimal supersymmetric Standard Model (MSSM) is a well-know
candidate for WIMP DM.

Since WIMPs are interacting with the ordinary matters, it is possible
to use these interactions to detect them directly. Experiments based on
such a strategy are called the DM direct detection experiments. These
experiments search for the scattering signals of WIMPs kicking off 
target nuclei on the earth by detecting the recoil energy transferred to
the nuclei. At present, the LUX experiment has the best sensitivity,
and provides a limit on the spin-independent WIMP--nucleon scattering
cross section as $\sigma_{\text{SI}}<7.6\times 
10^{-46}~\text{cm}^2$ at a mass of 33~GeV
\cite{Akerib:2013tjd}. Further, there exist several proposals with
ton-scale detectors, which will offer extremely improved sensitivities.

To study the properties of WIMPs based on the direct detection experiments,
it is necessary to evaluate the WIMP-nucleus scattering cross sections
accurately. The interactions of a WIMP with a nucleon, as well as a
nucleus, are generated through the couplings of the WIMP with quarks and
gluons. These couplings are described in terms of the parton-level
interactions, and contribute to the interactions with a nucleon through
non-perturbative QCD effects. An appropriate way to compute the
contribution is to take an effective theoretical approach. Here, the
parton-level interactions are expressed by the higher-dimensional
operators, and their contribution to the {\rm WIMP}-nucleon couplings is
computed by means of their nucleon matrix elements. See
Refs.~\cite{Drees:1993bu, Hisano:2010ct} for the treatments.  

Usually, the parton-level interactions are mediated by heavy
particles. Since the WIMPs are singlet with respect to the SU(3)$_{\rm
  C}\otimes$U(1)$_{\text{EM}}$ symmetry, the particles mediating the
WIMP-quark interactions should be also electrically neutral and color
singlet when they are exchanged in the $t$-channel, while they should
be charged and colored when they are exchanged in the $s$- or
$u$-channel. The extent of the significance of these contributions
highly depends on models. For example, in the case of the neutralino
DM in the MSSM, the $Z$ boson and the Higgs boson mediating processes
are classified into the former type, while the squark exchange is the
latter one. When squarks are extremely heavy, only the former
contribution is sizable. In the limit of pure gaugino or higgsino
case, on the other hand, the former contribution vanishes and thus the
latter may be dominant. Therefore, it is desirable to construct a
formalism to evaluate all of the contributions precisely enough on an
equal footing.

Recently, the LHC experiments give stringent limits on the masses of
new colored particles. The results may suggest that the colored
mediators which induce the couplings of WIMPs with quarks and gluon,
if they exist, should have masses much heavier than $\sim 100$~GeV. In this
case, the contribution of such a particle to the WIMP-nucleon
interaction receives sizable QCD corrections because of the mass
hierarchy and the large value of the strong coupling constant. This
motivates us to reformulate the calculation in the following way;
first, we obtain the effective theory which consists of the higher
dimensional operators of DM and quarks/gluons at the energy scale of
the mediator mass. At this point, we need to match the effective
theory to the full theory so that only the short-distance contribution
is to be included in the Wilson coefficients of the higher-dimensional
operators. Then, we evolve the operators by using the renormalization
group equations (RGEs) down to the scale at which the nucleon matrix
elements of the operators are evaluated. It is the prescription that
we will discuss in this paper. We will formulate a complete framework
to carry out the calculation to the leading order in the strong
coupling constant. In this formulation, all of the model dependence is
included into the Wilson coefficients of the effective operators, and
the rest part of the steps is independent of models. Therefore, the
method is quite suitable for generic computational codes such as {\tt
  micrOMEGAs} \cite{Belanger:2013oya} or {\tt DarkSUSY}
\cite{Gondolo:2004sc}.

In addition, we will discuss in detail the treatment for the twist-2
operators and show that we do not have to evolve the operators down to
the hadronic scale ($\sim 1$~GeV); we may evaluate their nucleon
matrix elements around the electroweak scale. This point is often
misunderstood, and we believe that our analysis clarifies the
confusion. Furthermore, we will show the significance of the
renormalization effects on the operators when the masses of the
colored mediators are much heavier than the electroweak scale.

This paper is organized as follows. In Sec.~\ref{sec:majoranaDM}, we
discuss our formulation in the case where DM is a Majorana fermion.
We present the effective operators for the Majorana fermion and evaluate the
matching conditions on their Wilson coefficients. The RGEs of the
operators are listed there as well. The results for {the real scalar 
and  vector boson DM} cases are also given in Sec.~\ref{sec:realscalarDM}
and Sec.~\ref{sec:realvectorDM}, respectively. Then, in
Sec.~\ref{sec:analysis}, we study the renormalization effects by using a
particular model. Also, we will discuss the
treatment of the twist-2 operators in the
section. Sec.~\ref{sec:conclusion} is devoted to conclusion. In
Appendix, we present the formulae for the one-loop
contribution to the scalar-type gluon operators in the case of scalar
DM, which as far as we know have not been given in the literature so
far. The result is useful when the colored mediators have masses around
the electroweak scale.

%%%%%%%%%%%%%%%%%%%%%%%%%%%%%%%%%%%
\section{Formalism: Majorana fermion  DM}
\label{sec:majoranaDM}
%%%%%%%%%%%%%%%%%%%%%%%%%%%%%%%%%%%

In this section, we give a formalism to evaluate the scattering cross
sections of a WIMP with a nucleon. The procedure described here
consists of the following steps. First, we construct the effective
theory for the WIMP, quarks, and gluons, by integrating out the
mediator particles. The effective interactions obtained here are
expressed in terms of the higher-dimensional operators. Then, we evolve
the Wilson coefficients of the effective operators according to the RGEs
down to the scale at which the nucleon matrix elements of the operators are
evaluated. Finally, we express the effective coupling of the WIMP 
with a nucleon in terms of the Wilson coefficients and the nucleon
matrix elements. The scattering cross sections are readily obtained from
the effective coupling. We will evaluate them to the leading order in
the strong coupling constant throughout this work. 

In this section,  we assume that the WIMP DM to be a Majorana fermion. The
real scalar boson DM and the real vector boson DM cases are discussed in
Sec.~\ref{sec:realscalarDM} and Sec.~\ref{sec:realvectorDM},
respectively. It should be noted that Dirac fermion and complex scalar
DM candidates are severely constrained by the DM direct detection
experiments because they in general couple to the vector current of quark
fields. Furthermore, this vector interaction is not renormalized, and
thus the conventional way of calculation is sufficient for this
contribution \cite{Jungman:1995df}. For these reasons, we do not
consider these cases in this paper.

%%%%%%%%%%%%%%%%%%%%%%%%%%%%%%%%%%%%
\subsection{Effective Lagrangian}
%%%%%%%%%%%%%%%%%%%%%%%%%%%%%%%%%%%

To begin with, let us write down the effective interactions of a
Majorana fermion, which is assumed to be a WIMP, with quarks and
gluon. The interactions are expressed in terms of the following
higher-dimensional operators \cite{Drees:1993bu}:
\begin{equation}
 {\cal L}_{\rm eff} = 
\sum_{p=q,g}C^p_S {\cal O}^p_S
+\sum_{i=1,2}\sum_{p=q,g}C^p_{T_i}{\cal O}^p_{T_i}
+\sum_{q} C^q_{AV}{\cal O}^q_{AV}
~,
\end{equation}
with
\begin{align}
 {\cal O}^q_S&\equiv
 \overline{\widetilde{\chi}^0}\widetilde{\chi}^0
 m_q\overline{q}q~,\nonumber \\ 
 {\cal O}^g_S&\equiv 
 \frac{\alpha_s}{\pi}
 \overline{\widetilde{\chi}^0}\widetilde{\chi}^0 
 G^A_{\mu\nu}G^{A\mu\nu}_{}~,\nonumber\\ 
 {\cal O}^p_{T_1}&\equiv \frac{1}{M}
 \overline{\widetilde{\chi}^0}i\partial^\mu \gamma^\nu
 \widetilde{\chi}^0 {\cal O}^p_{\mu\nu}~,\nonumber\\
 {\cal O}^p_{T_2}&\equiv \frac{1}{M^2}
 \overline{\widetilde{\chi}^0}i\partial^\mu i \partial^\nu
 \widetilde{\chi}^0 {\cal O}^p_{\mu\nu}~,\nonumber\\
 {\cal O}^q_{AV}&\equiv
 \overline{\widetilde{\chi}^0}\gamma_\mu^{}\gamma_5^{}
\widetilde{\chi}^0 \overline{q}\gamma^\mu_{}\gamma_5^{}
 q
~.
\end{align}
Here, we only keep the operators that remain sizable in the
non-relativistic limit. In addition, we have used the classical
equations of motion and the integration by parts to drop the redundant
operators \cite{Politzer:1980me, Arzt:1993gz}.   
$\widetilde{\chi}^0$, $q$, and $G^A_{\mu\nu}$ denote the Majorana
fermion, quarks ($q=u,d,s,c,b,t$), and the field strength tensor of
gluon field, respectively; $m_q$ are the masses of quarks; $M$ is the
mass of the WIMP; $\alpha_s\equiv g_s^2/(4\pi)$ is
the strong coupling constant, ${\cal O}^q_{\mu\nu}$ and
${\cal O}^g_{\mu\nu}$ are the twist-2 operators of quarks and gluon,
respectively, which are defined by\footnote{Notice that we have changed
the definition of ${\cal O}^g_{\mu\nu}$ by a factor of $-1$ from those in
Refs.~\cite{Drees:1993bu, Hisano:2010ct}. We follow the convention used
in Ref.~\cite{Buras:1979yt}. }
\begin{align}\label{twist2op}
 {\cal O}^q_{\mu\nu}&\equiv \frac{1}{2}\overline{q}i\biggl(
D_\mu^{}\gamma_\nu^{} +D_\nu^{}\gamma_\mu^{}-\frac{1}{2}g_{\mu\nu}^{}
\Slash{D}\biggr)q~,\nonumber \\
{\cal O}^g_{\mu\nu}&\equiv 
G^{A\rho}_{\mu} G^{A}_{\nu\rho}-\frac{1}{4}g_{\mu\nu}^{}
G^A_{\rho\sigma}G^{A\rho\sigma}~,
\end{align}
with $D_\mu$ the covariant derivatives. The effective operators are
defined at the mass scale of mediators, which is assumed to be well above
the mass of top quark. Generalization 
to other cases is straightforward; for instance, if such heavy particles
have masses similar to or lighter than the top mass, one should
integrate top quark as well so that the effective theoretical approach
is appropriate.\footnote{In Ref.~\cite{Gondolo:2013wwa}, such a situation is
discussed where the exchanged particle has a similar mass to the
$b$-quark mass. In this case, of course, $b$-quark (also top quark) should be
simultaneously integrated out when the effective theory is
formulated.\label{gondolo} }

Note that we include $\alpha_s/\pi$ to the
definition of the gluon scalar-type operator $\mathcal{O}^g_S$. We
discuss the meaning in the next subsection.

%%%%%%%%%%%%%%%%%%%%%%%%%%%%%%%%%%%%%%%%%%%%%%
\subsection{Nucleon matrix elements}
\label{sec:nucmat}
%%%%%%%%%%%%%%%%%%%%%%%%%%%%%%%%%%%%%%%

As discussed in Introduction, we need the nucleon matrix elements of
the effective operators to evaluate the WIMP-nucleon effective
coupling. These operators are classified into three types in terms of the
Lorentz transformation properties of the quark bilinear parts in the
operators; the scalar-type operators 
($\mathcal{O}^q_S$, $\mathcal{O}^g_S$), the axial-vector operator
($\mathcal{O}^q_{AV}$), and the twist-2-type operators
($\mathcal{O}^q_{T_i}$, $\mathcal{O}^g_{T_i}$). Since these operators do
not mix with each other under the renormalization group {(RG)} flow, we are allowed
to consider them separately.

%%%%%%%%%%%%%%%%%%%%%%%%%%%%%%%%%%%%%%%%%%%%%%%%
\begin{table}[t]
\caption{Mass fractions. These values are based on the lattice
 QCD simulations \cite{Young:2009zb, Oksuzian:2012rzb}. }
\label{table:massfraction}
 \begin{center}
\begin{tabular}{ll|ll}
\hline
\multicolumn{2}{c|}{Proton}&
\multicolumn{2}{c}{Neutron}\cr
\hline
\hline 
$f^{(p)}_{T_u}$& 0.019(5)&$f^{(n)}_{T_u}$&0.013(3)\cr
$f^{(p)}_{T_d}$& 0.027(6)&$f^{(n)}_{T_d}$& 0.040(9)\cr
$f^{(p)}_{T_s}$&0.009(22)&$f^{(n)}_{T_s}$& 0.009(22) \cr
\hline
\end{tabular}
\end{center}
\end{table}
%%%%%%%%%%%%%%%%%%%%%%%%%%%%%%%%%%%%%%%%%%%%%%%%%%%%%%%%%

As for the scalar-type quark operators $\mathcal{O}^q_S$, we use the
results from the lattice QCD simulations. The expectation values of
the scalar bilinear operators of light quarks between the nucleon states
at rest, $|N\rangle~(N=p,n)$, are parametrized as
\begin{equation}\label{eq:quarkscalar}
 f_{T_q}^{(N)}\equiv \langle N|m_q \bar{q}q|N\rangle/m_N ~,
\end{equation}
which are called the mass fractions. These values are shown in
Table~\ref{table:massfraction}. Here, $m_N$ is the nucleon mass. They
are taken from Ref.~\cite{Hisano:2012wm}, in which the mass fractions
are computed by using the results from Refs.~\cite{Young:2009zb,
Oksuzian:2012rzb}.

The nucleon matrix element of ${\cal O}^g_{S}$ is, on the other hand,
evaluated with the trace anomaly of the energy-momentum tensor
\cite{Shifman:1978zn}. For $N_f=3$ quark flavors, the trace of the
energy-momentum tensor in QCD is given as
\begin{equation}
 \Theta^\mu_{~\mu}=-\frac{9}{8}\frac{\alpha_s}{\pi} 
G^A_{\mu\nu}G^{A\mu\nu}_{}
  +\sum_{q=u,d,s}m_q\overline{q}q~,
\label{eq:traceanomaly}
\end{equation}
up to the leading order in $\alpha_s$. The relation beyond the leading
order in $\alpha_s$ is also readily obtained from the trace-anomaly
formula. By evaluating the operator \eqref{eq:traceanomaly} in the
nucleon states $|N\rangle$, from $ \langle N|
\Theta^\mu_{~\mu}|N\rangle=m_N$ we then obtain
\begin{equation}
 \langle N|\frac{\alpha_s}{\pi} G^A_{\mu\nu}G^{A\mu\nu}|N\rangle
=-\frac{8}{9} m_N f_{T_G}^{(N)}~,
\label{eq:gluonscalar}
\end{equation}
with $f_{T_G}^{(N)}\equiv 1-\sum_{q=u,d,s} f_{T_q}^{(N)}$. Notice that the
r.h.s. of Eq.~\eqref{eq:gluonscalar} is the order of the typical
hadronic scale, ${\cal O}(m_N)$. That is, although we include a factor
of $\alpha_s/\pi$ in the definition of ${\cal O}^g_{S}$, its nucleon matrix
element is not suppressed by $\alpha_s/\pi$. This is the
reason why we have defined ${\cal O}_{S}^g$ to contain $\alpha_s/\pi$.

%%%%%%%%%%%%%%%%%%%%%%%%%%%%%%%%%%%%%%%%%%%%%%%%
\begin{table}[t]
\caption{Second moments of the PDFs of proton evaluated at $\mu
 =m_Z$. We use the CJ12 next-to-leading order PDFs given by the
 CTEQ-Jefferson Lab collaboration \cite{Owens:2012bv}. }
\label{table:secondmoments}
 \begin{center}
\begin{tabular}{ll|ll}
\hline
\hline
$g(2)$ & 0.464(2) & &\\
$u(2)$ & 0.223(3) &$\bar{u}(2)$ & 0.036(2) \\
$d(2)$ & 0.118(3) &$\bar{d}(2)$ & 0.037(3) \\
$s(2)$ & 0.0258(4) &$\bar{s}(2)$ & 0.0258(4) \\
$c(2)$ & 0.0187(2) &$\bar{c}(2)$ & 0.0187(2) \\
$b(2)$ & 0.0117(1) &$\bar{b}(2)$ & 0.0117(1) \\
\hline
\hline
\end{tabular}
\end{center}
\end{table}
%%%%%%%%%%%%%%%%%%%%%%%%%%%%%%%%%%%%%%%%%%%%%%%%%%%%%%%%%

Next, we discuss the nucleon matrix elements of the twist-2
operators. They are given by the second moments of the parton
distribution functions (PDFs): 
\begin{align}
\langle N(p)\vert 
{\cal O}_{\mu\nu}^q
\vert N(p) \rangle 
&=
\frac{1}{m_N}\Bigl(p_{\mu}p_{\nu}-\frac{1}{4}m^2_N g_{\mu\nu}\Bigr)
(q(2;\mu)+\bar{q}(2;\mu))  ~,
%\label{eq:q2}
\\
\langle N(p) \vert 
{\cal O}_{\mu\nu}^g
\vert N(p) \rangle
& = -
\frac{1}{m_N}\Bigl(p_{\mu}p_{\nu}-\frac{1}{4}m^2_N g_{\mu\nu}\Bigr) 
g(2;\mu)  ~.
\label{eq:gt2mat}
\end{align}
with
\begin{align}
q(2;\mu) &= \int^1_0 dx~ x\ q(x,\mu)~\label{eq:q2},
\\
\bar{q}(2;\mu) &= \int^1_0 dx~ x\ \bar{q}(x,\mu)~\label{eq:qb2},
\\
g(2;\mu) &= \int^1_0 dx~ x\ g(x,\mu)~\label{eq:g2}.
\end{align}
Here $q(x,\mu)$, $\bar{q}(x,\mu)$ and $g(x,\mu)$ are the PDFs of quarks,
antiquarks and gluon at the factorization scale $\mu$,
respectively. These values are well measured at various energy scales,
contrary to the case of the scalar matrix elements. In
Table~\ref{table:secondmoments}, for example, we present the second moments
for proton at the scale of $\mu =m_Z$ with $m_Z$ the $Z$ boson mass. Here, we
use the CJ12 next-to-leading order PDFs given by the CTEQ-Jefferson Lab
collaboration \cite{Owens:2012bv}. Those for neutron are given with the
exchange of up 
and down quarks. As can be seen, the second moment for gluon 
$g(2;\mu)$ is of the same order of magnitude as those for quarks. As a
result, the nucleon matrix element of the gluon twist-2 tensor in
Eq.~\eqref{eq:gt2mat} is ${\cal O}(m_N)$. This justifies the definition of
${\cal O}^g_{T_i}$, where we have not included a factor of $\alpha_s/\pi$ in
this case. Our definition for the gluonic operators (${\cal O}^g_{S}$,
and ${\cal O}^g_{T_i}$) clarifies the order counting with respect to
$\alpha_s/\pi$.  

Finally, the nucleon matrix elements of the axial vector-type operators
are given by 
\begin{equation}
 \langle N \vert 
\bar{q}\gamma_{\mu}\gamma_5  q \vert N \rangle = 2 s_{\mu}\Delta q_N \ .
\end{equation}\label{eq:spinfraction}
with $\Delta q_N$ called the spin fractions and $s_{\mu}$ being the spin
of the nucleon. The values of the spin fractions are taken from
Ref.~\cite{Adams:1995ufa}; $\Delta u_p= 0.77$, $\Delta d_p= -0.49$, and
$\Delta s_p=-0.15$ for proton. Those of neutron are to be obtained by
exchanging the values of up and down quarks.

%%%%%%%%%%%%%%%%%%%%%%%%%%%%%%%%%%%%%%%%%%%
\subsection{Wilson coefficients}
%%%%%%%%%%%%%%%%%%%%%%%%%%%%%%%%%%%%%%%%%%%%

Next, we evaluate the Wilson coefficients of the effective operators by
integrating out heavy mediator particles. Here, we consider a generic
situation in which the interaction Lagrangian of the Majorana fermion with
quarks is given by
\begin{equation}
 {\cal L}_{\rm
  int}=\overline{q}(a_q^{}+b_q^{}\gamma_5^{})\widetilde{\chi}^0
  \widetilde {q} ~+~{\rm h.c.}~,
\label{eq:genlagmaj}
\end{equation}
where $\widetilde{q}$ denotes a heavy, colored scalar particle with its
mass represented by $M_{\widetilde{q}}$.

%%%%%%%FIGURE%%%%%%%%%%%%%%%%%%%%%%

\begin{figure}[t]
\begin{center}
 \includegraphics[height=40mm,clip]{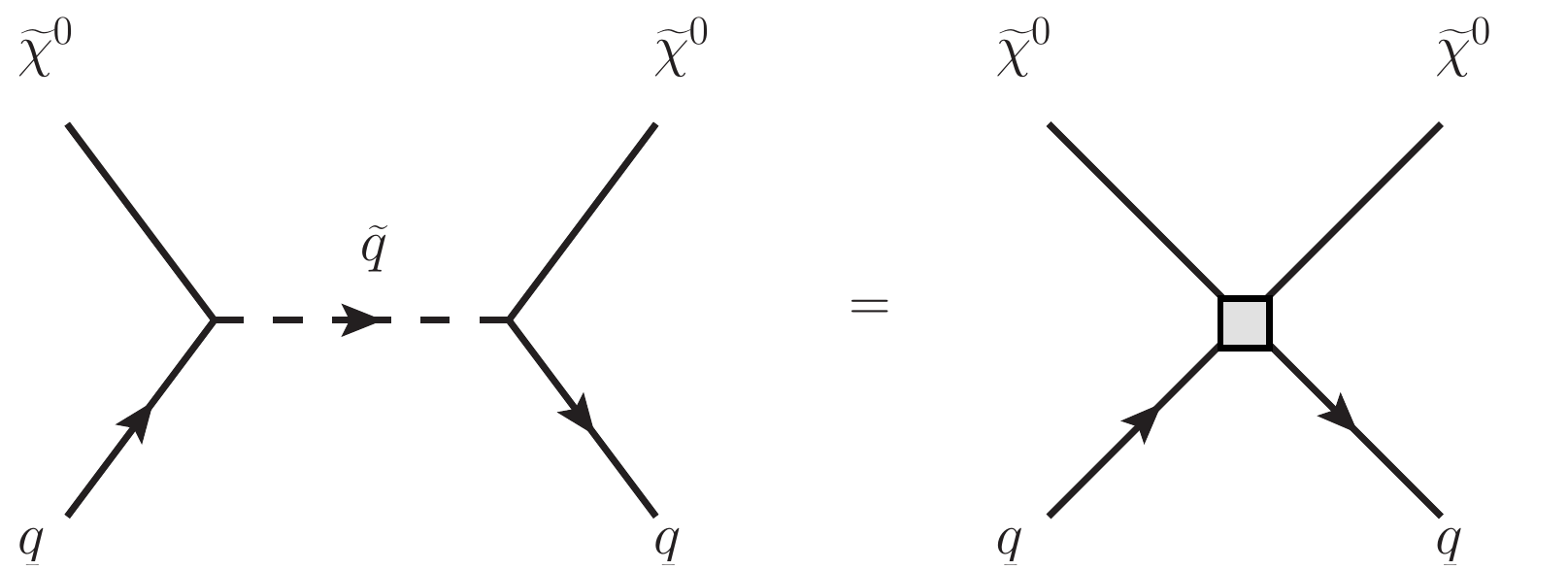}
\caption{Tree-level matching condition for Majorana fermion-quark effective
 interactions. Gray square represents the vertex for the quark effective
 operators. }
\label{fig:majtree}
\end{center}
\end{figure}

%%%%%%%%%%%%%%%%%%%%%%%%%%%%%%%%%

The interaction gives rise to the coupling of the Majorana fermion with
quarks via the tree-level exchange of the colored mediator. By
evaluating the diagram, we readily obtain the Wilson coefficients of the
effective operators containing quarks. The matching procedure is
illustrated in Fig.~\ref{fig:majtree}. Here, the gray square represents
the vertex for the WIMP-quark effective operators. As a result, we have
\begin{align}
 C^q_S(\mu_F)&=\frac{a_q^2+b_q^2}{8}
\frac{M}{(M_{\widetilde{q}}^2-M^2)^2} 
-\frac{a_q^2-b_q^2}{4m_q}\frac{1}{M_{\widetilde{q}}^2-M^2}~,
\nonumber \\
 C^q_{T_1}(\mu_F)&=
\frac{a_q^2+b_q^2}{2}\frac{M}{(M_{\widetilde{q}}^2-M^2)^2}~,\nonumber \\
 C^q_{T_2}(\mu_F)&=0~,\nonumber \\
 C^q_{AV}(\mu_F)&=
\frac{a_q^2+b_q^2}{4}\frac{1}{M_{\widetilde{q}}^2-M^2}~.
\end{align}
Here, $\mu_F$ denotes the factorization scale, which is taken to be
the mass scale of exchanged scalar particles. We have performed the
expansion in terms of the quark momenta; \textit{i.e.}, this calculation
is valid when $(p\cdot q)/(M_{\widetilde{q}}^2-M^2) \ll 1$ with
$p_\mu$ and $q_\mu$ being the momenta of DM and quarks, respectively
(See also footnote~\ref{gondolo}).\footnote{On the other hand, it turns out 
that the WIMP-nucleon scattering cross sections are considerably enhanced
when $M_{\widetilde{q}}-M < 100~\text{GeV}$ \cite{Hisano:2011um}.}

%%%%%%%FIGURE%%%%%%%%%%%%%%%%%%%%%%

\begin{figure}[t]
\begin{center}
 \includegraphics[height=70mm,clip]{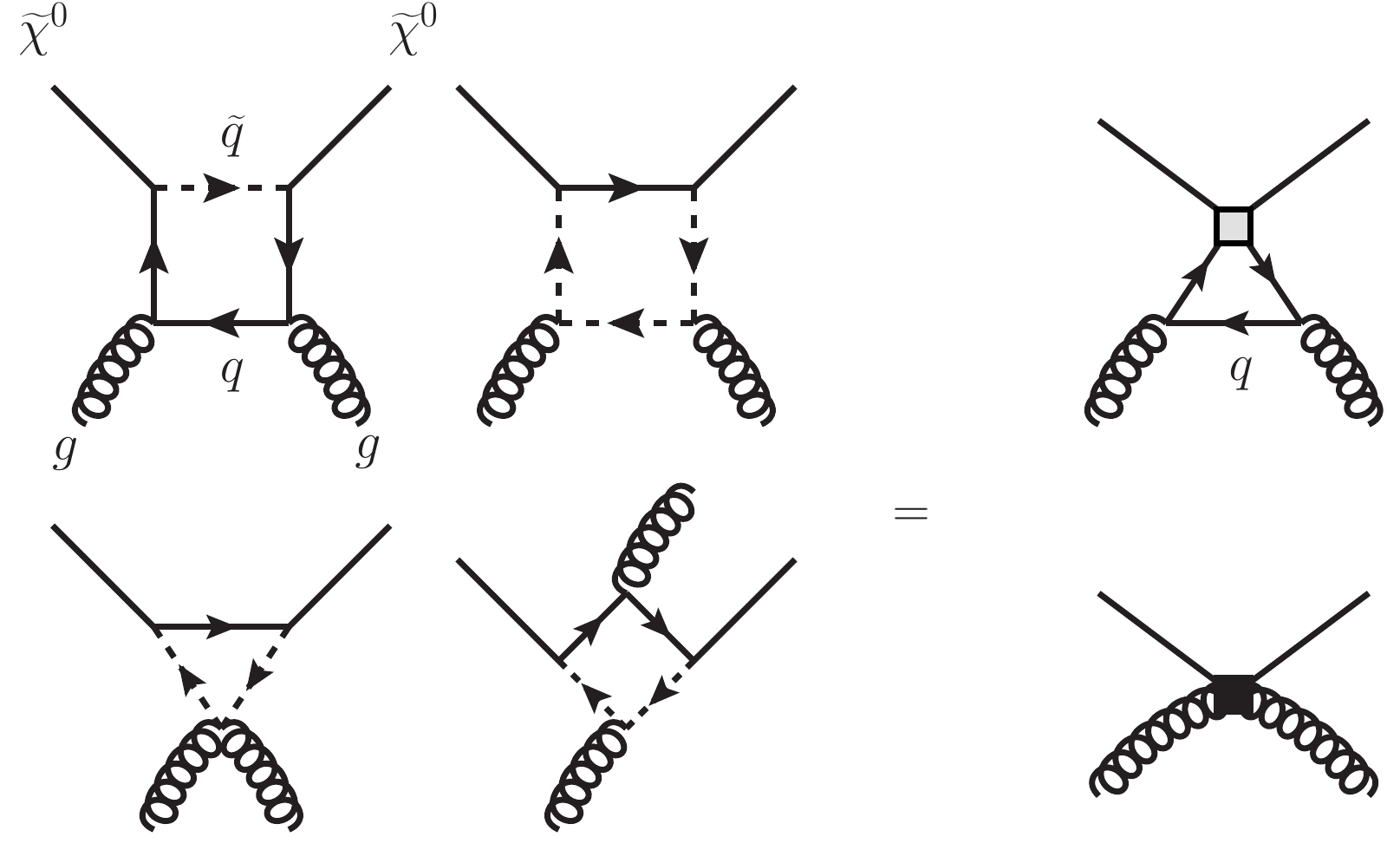}
\caption{One-loop matching condition for Majorana fermion-gluon effective
 interactions. Black square represents the vertex for the gluon
 effective operators. }
\label{fig:majglu}
\end{center}
\end{figure}

%%%%%%%%%%%%%%%%%%%%%%%%%%%%%%%%%

Next, we derive the matching condition for the WIMP-gluon effective
operators. At this point, we need to include only the short-distance
contribution to the Wilson coefficients \cite{Hisano:2010ct,
Hisano:2010fy}. This is achieved by the matching procedure shown in
Fig.~\ref{fig:majglu}. This reads 
\begin{align}
 C^g_S(\mu_F)&=-\sum_{q}\frac{M(a_q^2+b_q^2)}{96
 M_{\widetilde{q}}^2 (M_{\widetilde{q}}^2-M^2)}~,\nonumber \\
 C^g_{T_i}(\mu_F)&=0~.
\end{align}
Here, the summation is taken over all quark flavors. 
Note that the Wilson coefficient of the gluon scalar operator is
generated at the leading order in $\alpha_s$, \textit{i.e.},
$\mathcal{O}(\alpha_s^0)$, while those of the gluon twist-2 operators
vanish at this order \cite{Hisano:2010ct}. They are induced at
$\mathcal{O}(\alpha_s)$.

%%%%%%%%%%%%%%%%%%%%%%%%%%%%%%%%%%%%%%%%%%%%%
\subsection{Renormalization group equations}
\label{sec:RGEs}
%%%%%%%%%%%%%%%%%%%%%%%%%%%%%%%%%%%%%%%%%%%%%

The effective operators obtained above are evolved by means of
RGEs. In this section, we list the RGEs for the operators which we use
in the following analysis. In this paper, we only use the one-loop
RGEs since our main concern is to formulate the procedure for the
leading-order calculation in $\alpha_s$. For the sake of convenience,
however, we also mention some results that may be used for the
higher-order calculation.\footnote{See also Refs.~\cite{Hill:2011be,
    Hill:2014yxa} for relevant discussion.}

The one-loop beta function of the strong gauge coupling constant is
given by
\begin{align}
 \mu\frac{d\alpha_s}{d\mu}\equiv\beta
 (\alpha_s)=\frac{\alpha_s^2}{2\pi}
\biggl(-\frac{11}{3}N_c+\frac{2}{3}N_f\biggr)  ~,
\end{align}
where $N_c=3$ is the number of colors and $N_f$ denotes the number of quark
flavors in an effective theory. Higher-order contribution is found in
Refs.~\cite{vanRitbergen:1997va, Czakon:2004bu}.

Now we give the RGEs for the Wilson coefficients of the above
operators. First, we consider the RGEs for the scalar-type operators (${\cal O}^q_S, {\cal O}^g_S$). To
that end, notice that the quark mass operator is {RG} invariant in a mass-independent renormalization scheme like the
$\overline{\rm MS}$ scheme, {\it i.e.},
\begin{equation}
 \mu \frac{d}{d\mu}m_q\overline{q}q = 0~.
\end{equation}
Then, by differentiating the trace anomaly formula
\eqref{eq:traceanomaly}, we also find
\begin{equation}
 \mu \frac{d}{d\mu} \frac{\alpha_s}{\pi} G^A_{\mu\nu}G^{A\mu\nu}_{}=0~,
\end{equation}
as Eq.~\eqref{eq:traceanomaly} is an operator equation and thus
scale-invariant. 
Accordingly, the scalar-type operators are RG invariant at
$\mathcal{O}(\alpha_s)$. This is another reason why we include a factor
of $\alpha_s$ in the definition of $\mathcal{O}^g_S$. Beyond the leading
order, the scalar-type gluon operator runs and mixes with the
scalar-type quark
operators during the RG flow. The RGEs for the case are obtained again by
using the trace-anomaly formula with the use of the higher-order beta
function of the gauge coupling constant \cite{vanRitbergen:1997va,
Czakon:2004bu} and anomalous dimensions for quark masses
\cite{Chetyrkin:1997dh, Vermaseren:1997fq}.  

Next, we consider the RGEs for the twist-2 operators ($O^q_{T_i},
O^g_{T_i}$). The one-loop anomalous dimension matrix of the operators
is evaluated as \cite{Gross:1974cs}
\begin{equation}
 \mu \frac{d}{d\mu}(C^q_{T_i}, C^g_{T_i})=
(C^q_{T_i}, C^g_{T_i})~ \Gamma_T~,
\end{equation}
with $\Gamma_T$ a $(N_f+1)\times (N_f+1)$ matrix:
\begin{equation}
 \Gamma_T=\frac{\alpha_s}{4\pi}
\begin{pmatrix}
 \frac{16}{3}C_F&0&\cdots&0&\frac{4}{3}\\
 0 &\frac{16}{3}C_F&&\vdots&\vdots\\
\vdots&&\ddots&0&\vdots\\
0&\cdots&0&\frac{16}{3}C_F&\frac{4}{3}\\
\frac{16}{3}C_F&\cdots&\cdots&\frac{16}{3}C_F&\frac{4}{3}N_f
\end{pmatrix}
~,
\end{equation}
where $C_F=4/3$ is the quadratic Casimir invariant. Higher order RGEs
are found in Ref.~\cite{Vogt:2004mw}.

Finally, the RGE for the axial-vector interaction is readily obtained
since at the leading order the axial-vector current is conserved:
\begin{equation}
 \mu \frac{d}{d \mu}C^q_{AV} = 0~.
\end{equation}
For higher-order corrections, see Ref.~\cite{Larin:1993tq}.

%%%%%%%%%%%%%%%%%%%%%%%%%%%%%%%%%%%%%%%%%%%
\subsection{Quark threshold matching}
\label{sec:quarkthrmat}
%%%%%%%%%%%%%%%%%%%%%%%%%%%%%%%%%%%%%%%%%

During the RG flow, another matching procedure is required when one
goes across a quark threshold. For instance, around the $t$-quark mass
threshold $\mu_t \simeq m_t$, the Wilson coefficients are matched as 
\begin{align}
 C^q_{S}(\mu_t)|_{N_f=5}&= C^q_{S}(\mu_t)|_{N_f=6}~,\nonumber \\
C^g_{S}(\mu_t)|_{N_f=5}&=
-\frac{1}{12} \biggl[1+\frac{11}{4\pi}\alpha_s(\mu_t)\biggr]
C^t_{S}(\mu_t)|_{N_f=6}+ 
 C^g_{S}(\mu_t)|_{N_f=6}~,\nonumber \\
 C^q_{T_i}(\mu_t)|_{N_f=5}&= C^q_{
 T_i}(\mu_t)|_{N_f=6}~,\nonumber \\
 C^g_{T_i}(\mu_t)|_{N_f=5}&= 
C^g_{T_i}(\mu_t)|_{N_f=6}
~, \nonumber \\
 C^q_{AV}(\mu_t)|_{N_f=5}&=C^q_{AV}(\mu_t)|_{N_f=6}~,
\end{align}
with $q=u,d,s,c,b$. Notice that although we consider the leading-order
calculation, we include the next-to-leading order constant contribution
to $C^g_{S}$ here, since the effect is known to be large
\cite{Djouadi:2000ck}. Similar matching should be carried out at the
$b$- and $c$-quark threshold scales, if the coefficients are evolved
down below them. 

In addition, at the threshold scales, the higher-dimensional operators
suppressed by a power of the corresponding quark masses may also be
generated. For example, at a heavy quark mass threshold $m_Q$, the
$m_Q\overline{Q}Q$ operator gives rise to not only the scalar-type gluon
operator $-\alpha_s (m_Q)G^A_{\mu\nu}G^{A\mu\nu}/(12\pi)$, but also the
following dimension-six operators \cite{Cho:1994yu, Vecchi:2013iza}:
\begin{equation}
 -\frac{\alpha_s(m_Q)}{60\pi m_Q^2} (D^\nu G^A_{\nu\mu}) (D^\rho
  G^A_{\rho \mu})
-\frac{g_s\alpha_s(m_Q)}{720\pi m_Q^2}
f_{ABC} G^A_{\mu\nu} G^{B\mu\rho} G^C_{\nu\rho} ~,
\label{eq:dimsix}
\end{equation}
with $f_{ABC}$ the structure constant of SU(3)$_{\rm C}$. Among them,
those induced at the charm quark threshold may yield a significant
effect.  The naive dimensional analysis tells us that the higher
dimensional operator \eqref{eq:dimsix}  might give a correction up to
$\Lambda_{\rm QCD}^2/m_c^2\simeq 10$\% to the leading term, though the 
correction may be parametrically suppressed to a few \% by the prefactors of
those operators.

At present, there is no way to estimate the contribution of the
higher-dimensional operators more accurately. Thus, it should be
considered as a theoretical uncertainty of the computation. One way to
reduce the uncertainty is to use the nucleon matrix elements evaluated
above the charm threshold. As will be discussed in
Sec.~\ref{sec:analysis}, for the twist-2 operators, it is possible to
use the PDFs evaluated above the charm/bottom threshold. For the
scalar-type operators, on the other hand, the present lattice
simulations are not able to precisely evaluate the charm-quark matrix
element \cite{Dinter:2012tt}. Future simulations may compute it with
sufficient accuracy and help to reduce the theoretical uncertainty.

%%%%%%%%%%%%%%%%%%%%%%%%%%%%%%%%%%%%%%%
\subsection{Scattering cross sections}
%%%%%%%%%%%%%%%%%%%%%%%%%%%%%%%%%%%%%%%

Finally, we obtain the effective coupling of the Majorana fermion with a
nucleon. For the spin-independent coupling, we have
\begin{equation}
 \mathcal{L}^{(N)}_{\text{SI}}=f_N\overline{\widetilde{\chi}^0}
\widetilde{\chi}^0 \overline{N}N~,
\end{equation}
where $N$ denotes the nucleon field and 
\begin{align}
 f_N/m_N&= \sum_{q=u,d,s}C^q_S(\mu_{\text{had}}) f^{(N)}_{T_q} 
-\frac{8}{9}C^g_S(\mu_{\text{had}}) f^{(N)}_{T_G}
\nonumber \\
&+\frac{3}{4}\sum_{q}^{N_f}
 \sum_{i=1,2}C^q_{T_i}(\mu)[q(2;\mu)+\overline{q}(2;\mu)]
-\frac{3}{4}\sum_{i=1,2}C^g_{T_i}(\mu)g(2;\mu)~,
\end{align}
where $\mu_{\text{had}}$ is the hadron scale usually taken to be around
1~GeV. 
For the contribution of the twist-2 operators, the summation runs over the
number of the active quark flavors in the effective theory at the energy
scale of $\mu$ where their nucleon matrix elements are evaluated. As
will be shown in Sec.~\ref{sec:analysis}, the scale $\mu$ does not need
to be taken at the hadronic scale; it is allowed to be set around the
electroweak scale as long as the PDFs of the scale are known.

The
spin-dependent effective coupling is, on the other hand, given by
\begin{equation}
 \mathcal{L}^{(N)}_{\text{SD}}=a_N\overline{\widetilde{\chi}^0}
\gamma^\mu \gamma_5
\widetilde{\chi}^0 \overline{N}\gamma_\mu \gamma_5N~,
\end{equation}
with
\begin{equation}
 a_N=\sum_{q=u,d,s}C^q_{AV}\Delta q_N~.
\end{equation}

By using the effective couplings, we finally obtain the scattering cross
section of the Majorana fermion with a target nucleus as follows:
\begin{equation}
 \sigma =\frac{4}{\pi}\biggl(\frac{MM_T}{M+M_T}\biggr)^2
\biggl[
|n_pf_p+n_nf_n|^2
+4\frac{J+1}{J}
|a_p\langle s_p\rangle
+a_n\langle s_n\rangle|^2
\biggr]~,
\end{equation}
where $M_T$ is the mass of the target nucleus; $n_p$ and $n_n$ are the
numbers of protons and neutrons in the nucleus, respectively; $J$ is the
total spin of the nucleus, and $\langle s_N\rangle$ is the expectation
value of the spin of a nucleon in the target. Here, we calculate
WIMP-nucleus scattering cross sections in the limit of zero momentum
transfer, for which each WIMP-nucleon scattering amplitude adds up
coherently \cite{Jungman:1995df}.

%%%%%%%%%%%%%%%%%%%%%%%%%%%%%%%%%%%%%%
\section{Formalism: real scalar boson  DM}
\label{sec:realscalarDM}
%%%%%%%%%%%%%%%%%%%%%%%%%%%%%%%%%%%%%

Next we briefly show the results for the case of  {real scalar boson DM}. We
may use a similar procedure to that given in the previous section to
formulate effective theories for the WIMP. 

%%%%%%%%%%%%%%%%%%%%%%%%%%%%%%%%%%%%%%%%%%%
\subsection{Effective Lagrangian}
%%%%%%%%%%%%%%%%%%%%%%%%%%%%%%%%%%%%%%%%%

The effective interactions of the real scalar $\phi$ with quarks and
gluon are expressed by 
\begin{equation}
\mathcal{L}_{\rm eff}
=\sum_{p=q,g}C^p_S\mathcal{O}^p_S
+\sum_{p=q,g}C^p_{T_2}\mathcal{O}^p_{T_2}~,
\label{LeffscalarDM}
\end{equation} 
with
\begin{align}
\mathcal{O}^q_S&\equiv  \phi^2 m_q\bar{q}q~,\nonumber\\
\mathcal{O}^g_S&\equiv \frac{\alpha_s}{\pi}
\phi^2  G^{A\mu\nu}G^A_{\mu\nu}~,\nonumber\\
\mathcal{O}^q_{T_2}&\equiv\frac{1}{M^2}\phi i \partial^\mu i
 \partial^\nu \phi \mathcal{O}^q_{\mu\nu}~,\nonumber\\ 
\mathcal{O}^g_{T_2}&\equiv\frac{1}{M^2}\phi i\partial^\mu i\partial^\nu
 \phi \mathcal{O}^g_{\mu\nu}~.
\end{align}
Note that there is no spin-dependent interactions in the case of 
scalar boson DM.

%%%%%%%%%%%%%%%%%%%%%%%%%%%%%%%%%%%%%%%%%%%
\subsection{Wilson coefficients}
%%%%%%%%%%%%%%%%%%%%%%%%%%%%%%%%%%%%%%%%%

%%%%%%%FIGURE%%%%%%%%%%%%%%%%%%%%%%

\begin{figure}[t]
\begin{center}
 \includegraphics[height=40mm,clip]{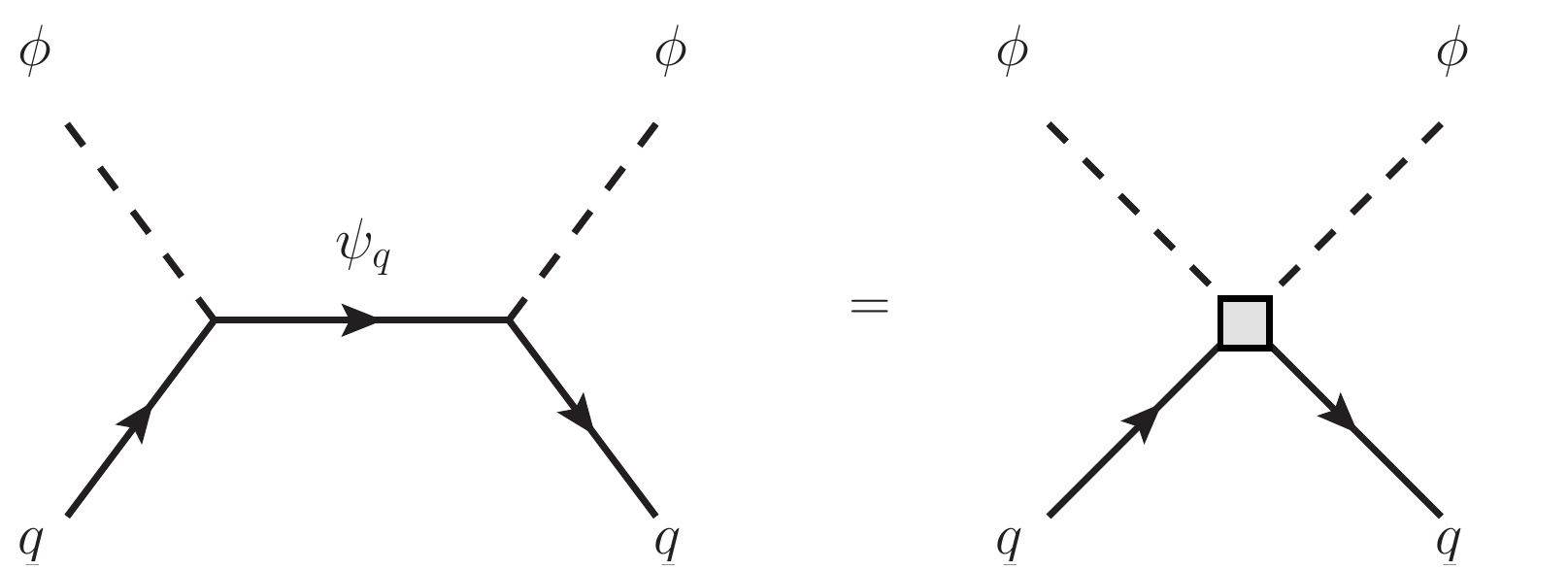}
\caption{Tree-level matching condition for scalar boson-quark effective
 interactions. Gray square represents the vertex for the quark effective
 operators. }
\label{fig:scatree}
\end{center}
\end{figure}

%%%%%%%%%%%%%%%%%%%%%%%%%%%%%%%%%

\begin{figure}[t]
\begin{center}
 \includegraphics[height=70mm,clip]{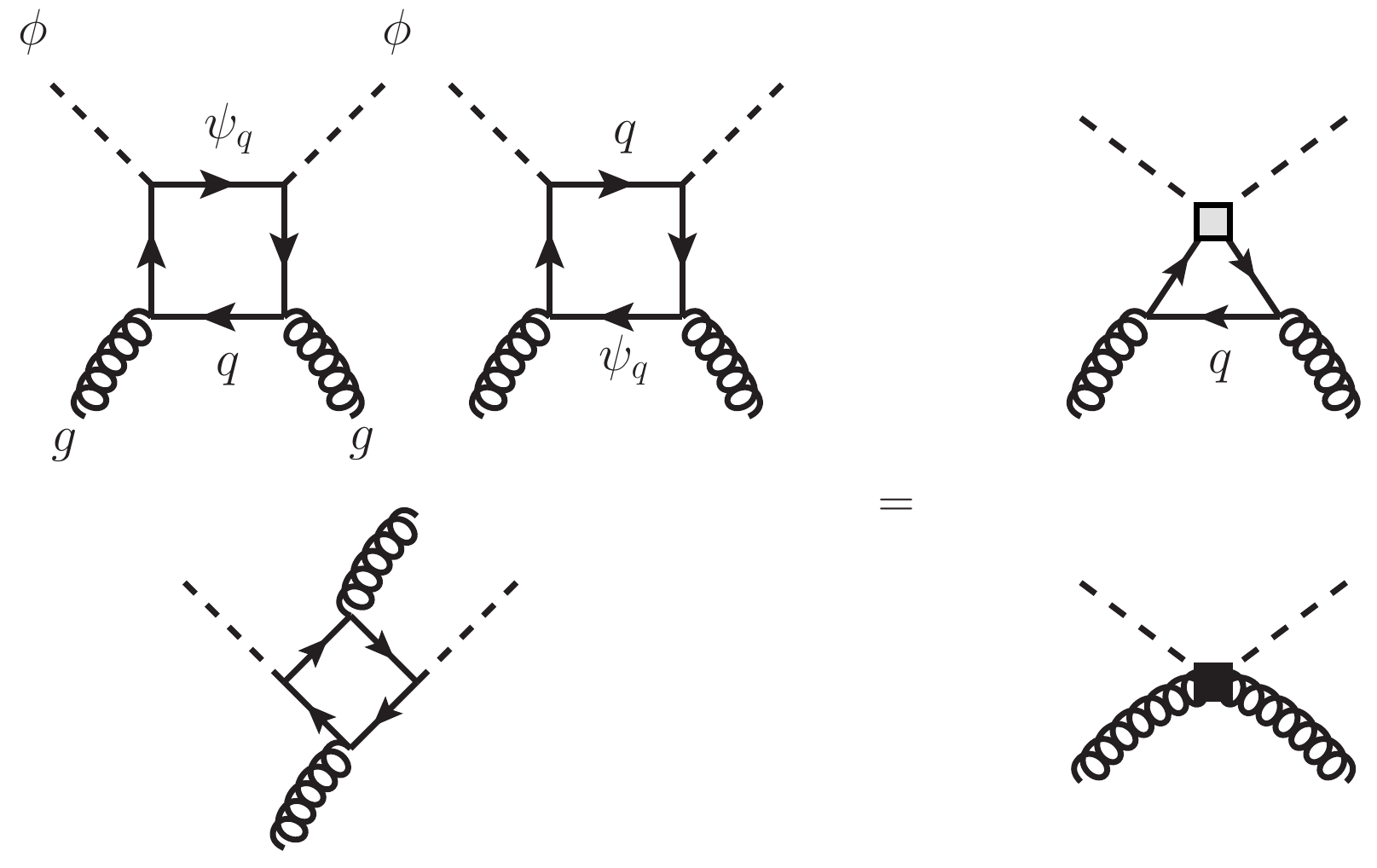}
\caption{One-loop matching condition for scalar boson-gluon effective
 interactions. Black square represents the vertex for the gluon
 effective operators. }
\label{fig:scaglu}
\end{center}
\end{figure}

We next discuss the matching condition for the Wilson coefficients of
the above operators in a theory where the interactions of the scalar boson
with quarks are given by
\begin{equation}
\mathcal{L}=\overline{\psi}_q(a_q+b_q\gamma_5 )q  \phi+ \text{h.c.}~,
\label{full:scalar}
\end{equation}
where $\psi_q$ denotes a colored fermion with a mass of
$M_{\psi_q}$. Then, with the tree-level matching procedure for the
WIMP-quark interactions illustrated in Fig.~\ref{fig:scatree}, we obtain 
\begin{align}
C^q_S(\mu_F)&=\frac{a^2_q+b^2_q}{2}
 \frac{2M^2_{\psi_q}-M^2}{(M^2_{\psi_q}-M^2)^2}
+
\frac{a^2_q-b^2_q}{m_q}\frac{M_{\psi_q}}{M^2_{\psi_q}-M^2}
~,
\label{CqSscalar}\\[3pt]
C^q_{T_2}(\mu_F)&=\frac{2(a^2_q+b^2_q) M^2}{(M^2_{\psi_q}-M^2)^2}~.
\end{align}
Again, the calculation is valid only when the mass difference between
the heavy mediator particle and the real scalar boson is much larger than
the energy of external quarks. The matching condition for the gluon
operators is, on the other hand, obtained through the procedure shown in
Fig.~\ref{fig:scaglu}. We have
\begin{align}
C^g_S(\mu_F)&=
 \sum_{q}\frac{a^2_q+b^2_q}{12(M^2_{\psi_q}-M^2)}~,\\ 
C^g_{T_2}(\mu_F)&= 0~.
\label{CgSscalar}
\end{align} 
Further, in Appendix, we give a result for the
loop-computation of the one-loop diagrams in the left-hand side in
Fig.~\ref{fig:scaglu}, since as far as we know there has been no such a
calculation in the literature. The result is useful for the cases where
the masses of the colored particles are not so heavy compared with some
of the quark masses.

%%%%%%%FIGURE%%%%%%%%%%%%%%%%%%%%%%

%%%%%%%%%%%%%%%%%%%%%%%%%%%%%%%%%

%%%%%%%%%%%%%%%%%%%%%%%%%%%%%%%%%%%%%%%%%%%%%%
\subsection{Scattering cross sections}
%%%%%%%%%%%%%%%%%%%%%%%%%%%%%%%%%%%%%%%

We now ready to evaluate the scattering cross section of {the real scalar boson}
with a target nucleus. The spin-independent coupling of the real
scalar boson with a nucleon defined by 
\begin{eqnarray}
\mathcal{L}^{(N)}_{\text{SI}}=f_N\phi^2 \overline{N}N~,
\end{eqnarray}
is evaluated as
\begin{align}
 f_N/m_N&= \sum_{q=u,d,s}C^q_S(\mu_{\text{had}}) f^{(N)}_{T_q} 
-\frac{8}{9}C^g_S(\mu_{\text{had}}) f^{(N)}_{T_G}
\nonumber \\
&+\frac{3}{4}\sum_{q}^{N_f}
C^q_{T_2}(\mu)[q(2;\mu)+\overline{q}(2;\mu)]
-\frac{3}{4}C^g_{T_2}(\mu)g(2;\mu)~.
\end{align}
In the scalar boson case, there is no spin-dependent coupling with a
nucleon.  By using the effective coupling, we calculate the scattering
cross section of the real scalar boson with a target nucleus as follows: 
\begin{equation}
 \sigma =\frac{1}{\pi}\biggl(\frac{M_T}{M+M_T}\biggr)^2
|n_pf_p+n_nf_n|^2
~.
\end{equation}

%%%%%%%%%%%%%%%%%%%%%%%%%%%%%%%%%
\section{Formalism: real vector boson DM}
\label{sec:realvectorDM}
%%%%%%%%%%%%%%%%%%%%%%%%%%%%%%%%%%

Finally, we consider {real vector boson DM}. For previous calculation, see
Ref.~\cite{Hisano:2010yh} and references therein.

%%%%%%%%%%%%%%%%%%%%%%%%%%%%%%%%%%%%%%%%%%%
\subsection{Effective Lagrangian}
%%%%%%%%%%%%%%%%%%%%%%%%%%%%%%%%%%%%%%%%%

The effective interactions of the real vector boson $B_\mu$ with quarks and
gluon are written as
\begin{equation}
\mathcal{L}_{\rm eff}
=\sum_{p=q,g}C^p_S\mathcal{O}^p_S
+\sum_{p=q,g}C^p_{T_2}\mathcal{O}^p_{T_2}
+\sum_{q}C^q_{AV}\mathcal{O}^q_{AV}~,
\label{LeffvectorDM}
\end{equation} 
with
\begin{align}
\mathcal{O}^q_S&\equiv  B^\mu B_\mu m_q\bar{q}q~,\nonumber\\
\mathcal{O}^g_S&\equiv \frac{\alpha_s}{\pi}B^\rho B_\rho 
 G^{A\mu\nu}G^A_{\mu\nu}~,\nonumber\\ 
\mathcal{O}^q_{T_2}&\equiv\frac{1}{M^2}B^\rho i \partial^\mu i
 \partial^\nu B_\rho \mathcal{O}^q_{\mu\nu}~,\label{WCs}\nonumber\\ 
\mathcal{O}^g_{T_2}&\equiv\frac{1}{M^2}B^\rho i\partial^\mu
 i\partial^\nu B_\rho \mathcal{O}^g_{\mu\nu}~,\nonumber\\ 
\mathcal{O}^q_{AV}&\equiv\frac{1}{M}\epsilon_{\mu\nu\rho\sigma}B^\mu
 i\partial^\nu B^\rho \bar{q}\gamma^\sigma\gamma_5 q~,
\end{align}
where $\epsilon^{\mu\nu\rho\sigma}$ is the totally antisymmetric tensor
with $\epsilon^{0123}\equiv +1$. Here, the vector boson field is supposed
to satisfy the on-shell condition $(\square +M^2)B_\mu =0$ and
$\partial_\mu B^\mu =0$.

%%%%%%%%%%%%%%%%%%%%%%%%%%%%%%%%%%%%%%%%%%%
\subsection{Wilson coefficients}
%%%%%%%%%%%%%%%%%%%%%%%%%%%%%%%%%%%%%%%%%%%

%%%%%%%FIGURE%%%%%%%%%%%%%%%%%%%%%%
\begin{figure}[t]
 \begin{center}
  \includegraphics[height=40mm]{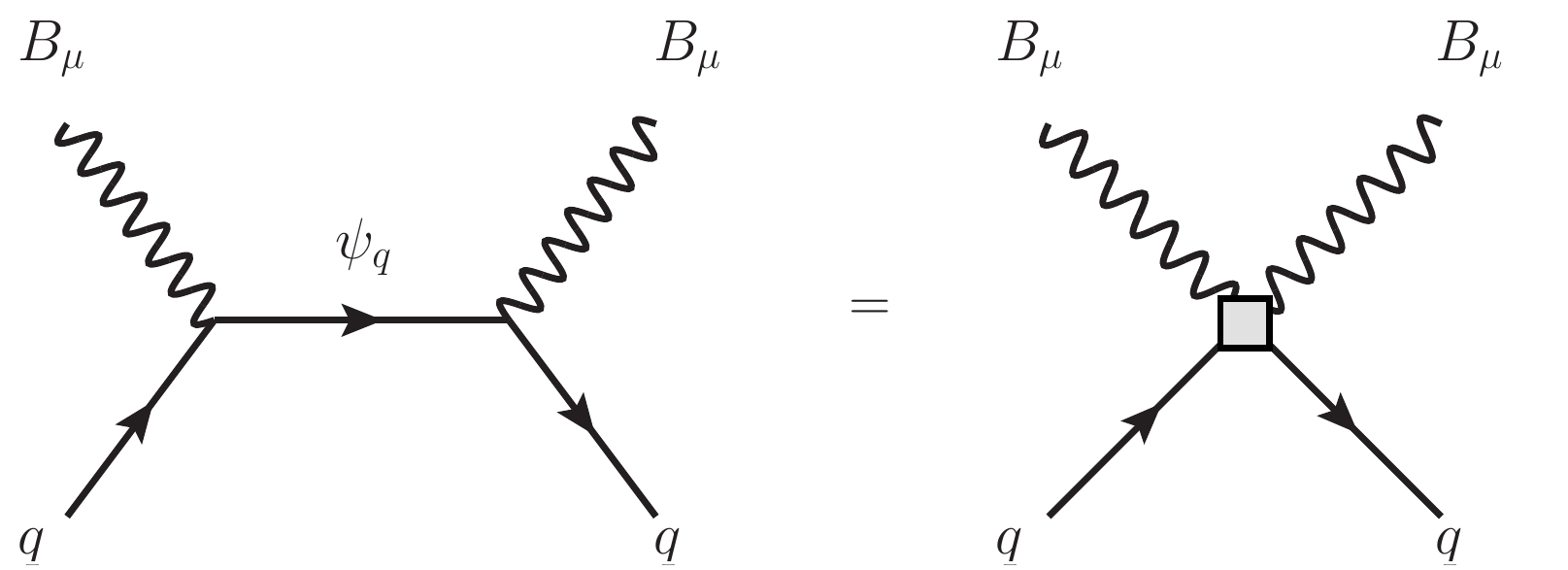}
 \end{center}
 \caption{Tree-level matching condition for vector boson-quark effective
 interactions. Gray square represents the vertex for the quark effective
 operators. }
 \label{figBBqq}
\end{figure}
%%%%%%%FIGURE%%%%%%%%%%%%%%%%%%%%%%

%%%%%%%FIGURE%%%%%%%%%%%%%%%%%%%%%%

\begin{figure}[t]
\begin{center}
 \includegraphics[height=70mm,clip]{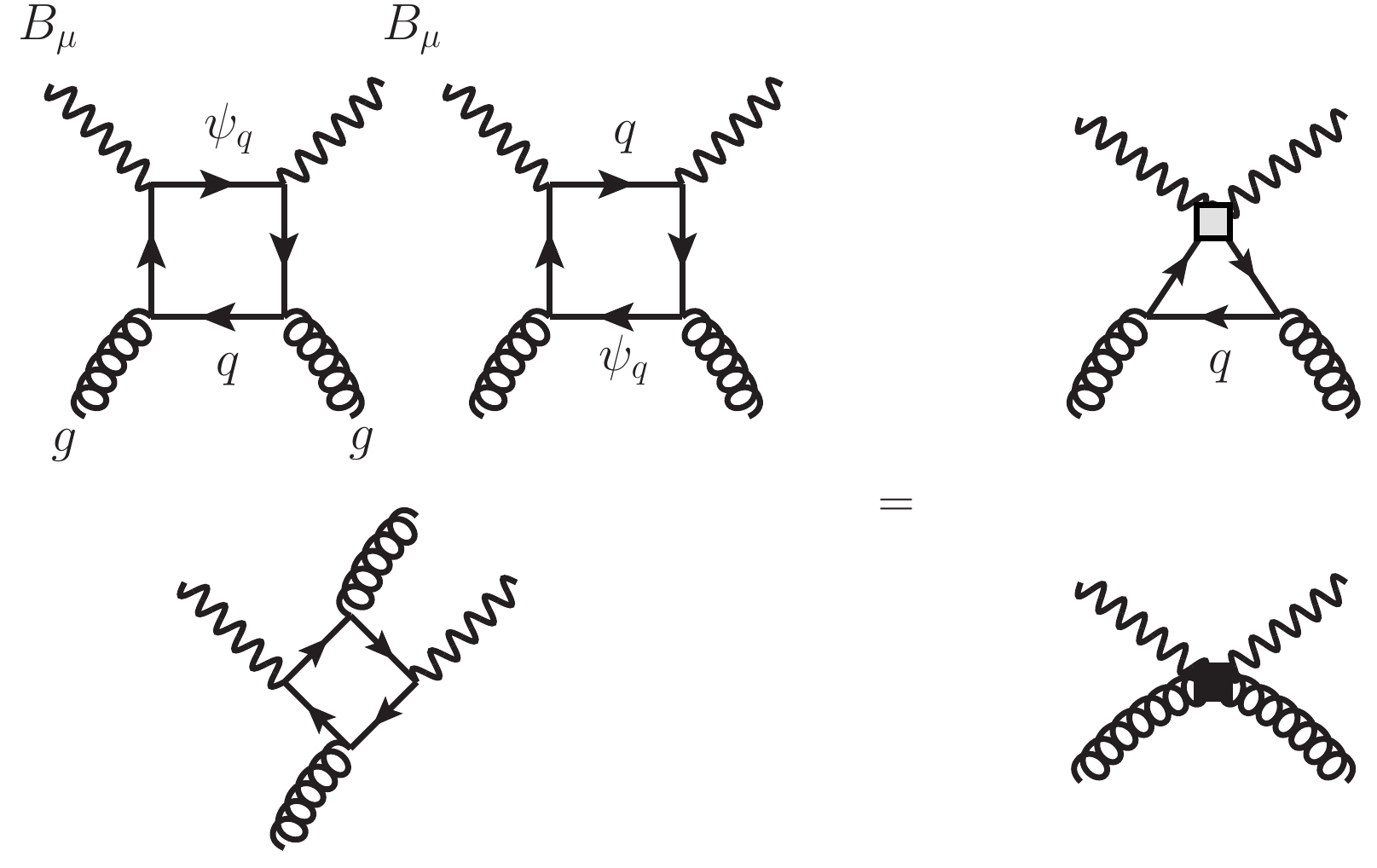}
\caption{One-loop matching condition for vector boson-gluon effective
 interactions. Black square represents the vertex for the gluon
 effective operators. }
\label{figBBGG}
\end{center}
\end{figure}

%%%%%%%%%%%%%%%%%%%%%%%%%%%%%%%%%

Let us evaluate the Wilson coefficients of the above operators in the
presence of a fermionic colored particle $\psi_q$ coupling to the WIMP and
quarks through the interactions:
\begin{eqnarray}
\mathcal{L}=\overline{\psi}_q
(a_q\gamma^\mu+b_q\gamma^\mu\gamma_5 )q  B_\mu+\text{h.c.}~.
\end{eqnarray}
In this case, Fig.~\ref{figBBqq} yields the matching condition for
the WIMP-quark effective couplings as
\begin{align}
C^q_S(\mu_F)&=
-(a^2_q+b^2_q)\frac{M^2_{\psi_q}}{2(M^2_{\psi_q}-M^2)^2} +
\frac{a^2_q-b^2_q}{m_q}\frac{M_{\psi_q}}{M^2_{\psi_q}-M^2}
~,\label{fq}\\[3pt] 
C^q_{T_2}(\mu_F)&=-\frac{2(a^2_q+b^2_q)M^2}{(M^2_{\psi_q}-M^2)^2}~,
\label{gq}\\[3pt]
C^q_{AV}(\mu_F)&=\frac{i(a^2_q+b^2_q)M}{M^2_{\psi_q}-M^2}~.\label{dq}
\end{align}
As for the gluon contribution, Fig.~\ref{figBBGG} reads
\begin{align}
C^g_S(\mu_F)&=
 \sum_{q}\frac{a^2_q+b^2_q}{12(M^2_{\psi_q}-M^2)}~,\\[3pt]
C^g_{T_2}(\mu_F)&=0~.
\end{align}

%%%%%%%%%%%%%%%%%%%%%%%%%%%%%%%%%%%%%%%%%%%%%%
\subsection{Scattering cross sections}
%%%%%%%%%%%%%%%%%%%%%%%%%%%%%%%%%%%%%%%

By using the results obtained above, we now evaluate the
spin-independent WIMP-nucleon coupling. With the definition
\begin{eqnarray}
\mathcal{L}^{(N)}_{\text{SI}}=f_NB_\mu B^\mu \overline{N}N~,
\end{eqnarray}
we have
\begin{align}
 f_N/m_N&= \sum_{q=u,d,s}C^q_S(\mu_{\text{had}}) f^{(N)}_{T_q} 
-\frac{8}{9}C^g_S(\mu_{\text{had}}) f^{(N)}_{T_G}
\nonumber \\
&+\frac{3}{4}\sum_{q}^{N_f}
C^q_{T_2}(\mu)[q(2;\mu)+\overline{q}(2;\mu)]
-\frac{3}{4}C^g_{T_2}(\mu)g(2;\mu)~.
\end{align}
On the other hand, the spin-dependent effective coupling is given by
\begin{eqnarray}
\mathcal{L}^{(N)}_{\text{SD}}=\frac{a_N}{M}\epsilon_{\mu\nu\rho\sigma}B^\mu
 i\partial^\nu B^\rho \overline{N}\gamma^\sigma \gamma_5 N~, 
\end{eqnarray}
where
\begin{equation}
 a_N=\sum_{q=u,d,s}C^q_{AV}\Delta q_N~.
\end{equation}
With these effective couplings, we eventually get the scattering cross
section of the real vector boson with a target nucleus as
\begin{equation}
 \sigma =\frac{1}{\pi}\biggl(\frac{M_T}{M+M_T}\biggr)^2
\biggl[
|n_pf_p+n_nf_n|^2
+\frac{8}{3}\frac{J+1}{J}
|a_p\langle s_p\rangle
+a_n\langle s_n\rangle |^2
\biggr]~.
\end{equation}

%%%%%%%%%%%%%%%%%%%%%%%%%%%%%%%%%%%%%%
\section{Analysis}
\label{sec:analysis}
%%%%%%%%%%%%%%%%%%%%%%%%%%%%%%%%%%%%%

Now we apply our formulation to a concrete DM model, and discuss the
renormalization effects on the calculation. Here, we consider a Majorana
fermion interacting with only the third generation right-handed quarks with a
unit coupling constant; \textit{i.e.}, $a_q = b_q = 0$ for $q=u,d,s,c$
and $a_q = b_q =1/2$ for $q=b,t$ in Eq.~\eqref{eq:genlagmaj}. Since the
new colored scalar particles introduced to the scenario only couple to
the third generation quarks, the LHC constraints on them are less severe
compared with those interacting with the first two generation quarks. 
This model is a simplified model for a system composed of a
neutralino DM and a pair of right-handed stop and sbottom in the MSSM,
though the couplings between them are different from above assumption.

%%%%%%%FIGURE%%%%%%%%%%%%%%%%%%%%%%

\begin{figure}[t]
\begin{center}
 \includegraphics[height=70mm,clip]{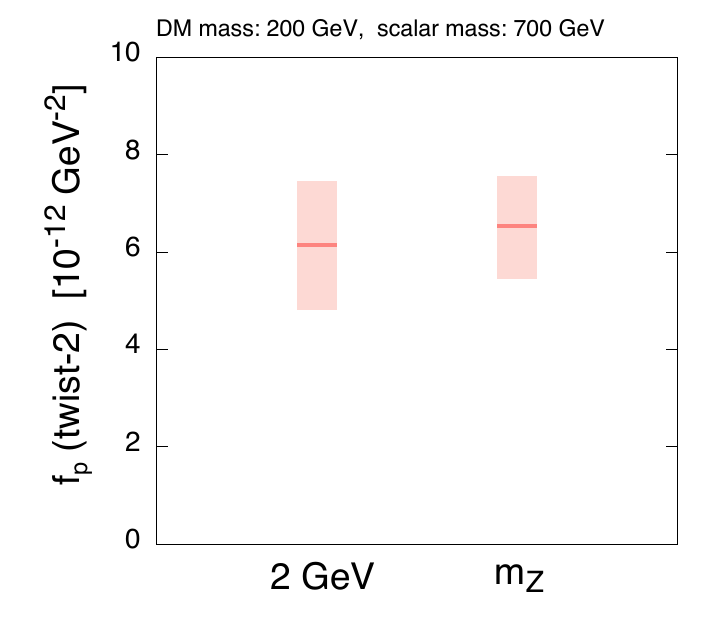}
 \caption{Comparison of  twist-2 contributions to  WIMP-proton
   effective coupling calculated with PDFs obtained at $\mu=2$~GeV and
   $m_Z$. Here, we assume the WIMP is a Majorana fermion coupled with
   right-handed $t$- and $b$-quarks (see text), and we take $M=200$~GeV and
   $M_{\widetilde{q}}= 700$~GeV. Red (light-pink) bar denotes the
   uncertainty coming from the PDF input (perturbation in $\alpha_s$).
 }
\label{fig:twist2mz2gev}
\end{center}
\end{figure}

%%%%%%%%%%%%%%%%%%%%%%%%%%%%%%%%%

By using the model, we first discuss the scale at which we evaluate the
nucleon matrix elements of the twist-2 operators. As mentioned in
Sec.~\ref{sec:nucmat},  {the 
twist-2 operators are not mixed with the scalar-type and axial-vector operators so that we may take a scale for the matrix elements of the twist-2 operators which is different from that of the other ones, and 
the matrix elements are obtained in a wide
range of energy scales.} Thus, it is important to determine which scale
is appropriate for the calculation of the twist-2 contribution. In
Fig.~\ref{fig:twist2mz2gev}, we compare the twist-2 contributions to the
WIMP-proton effective coupling,
\begin{equation}
  f_p~\text{(twist-2)}/m_p = \frac{3}{4}\sum_{q}^{N_f}
 \sum_{i=1,2}C^q_{T_i}(\mu)[q(2;\mu)+\overline{q}(2;\mu)]
-\frac{3}{4}\sum_{i=1,2}C^g_{T_i}(\mu)g(2;\mu)~,
\end{equation}
evaluated with the PDFs obtained at $\mu=2$~GeV and $m_Z$. Here, we take
$M=200$~GeV and $M_{\widetilde{q}}= 700$~GeV. The red (light-pink) bar
denotes the uncertainty coming from the PDF input (perturbation in $\alpha_s$). 
For the estimation of the uncertainty from the PDF error, we follow the
method described in Ref.~\cite{Owens:2012bv} with the $\chi^2$ tolerance
$T$ taken to be $T=10$. The uncertainty caused by the neglect of the
higher-order contribution in $\alpha_s$ is evaluated by varying the
input and quark-mass threshold scales by a factor of two, \textit{i.e.},
$M_{\widetilde{q}}/2\leq\mu_F \leq 2M_{\widetilde{q}}$, $m_t/2 \leq
\mu_t \leq 2m_t$, and so on. It turns out that both calculations
predict similar values for the twist-2 contribution, though the
theoretical error in the case of $\mu =2$~GeV is {a little bit} larger than that with
$\mu =m_Z$. If one sets $\mu = 1$~GeV, we expect that the error becomes
much larger due to the charm-quark threshold effects since in the
low-energy region the strong coupling constant rapidly grows up. In
addition, the higher-dimensional operators suppressed by a power of the
quark masses may give significant contribution if the scale $\mu$ is
taken to be at a low-energy scale as discussed in
Sec.~\ref{sec:quarkthrmat}, which also contribute to the theoretical
uncertainty. For these reasons, we conclude that it is appropriate to
set the PDF scale $\mu$ in a high-scale region, not the hadronic
scale. In the following calculation, we take the scale to be $\mu =
m_Z$.

Next, we show the renormalization effects on the WIMP-nucleon
scattering. As discussed in Sec.~\ref{sec:RGEs}, the twist-2 operators
receive the renormalization effects. The effects are expected to be
significant when the input scale $\mu_F$, {\it i.e.}, the typical mass
scale of colored mediators, is much higher than the electroweak scale.

%%%%%%%%%%%%%% FIGURE %%%%%%%%%%%%%%%%%%%%%%%%%%%%%%%%%%%%
\begin{figure}[t!]
\begin{center}
\subfigure[Effective coupling $f_p$]
 {\includegraphics[clip, width = 0.45 \textwidth]{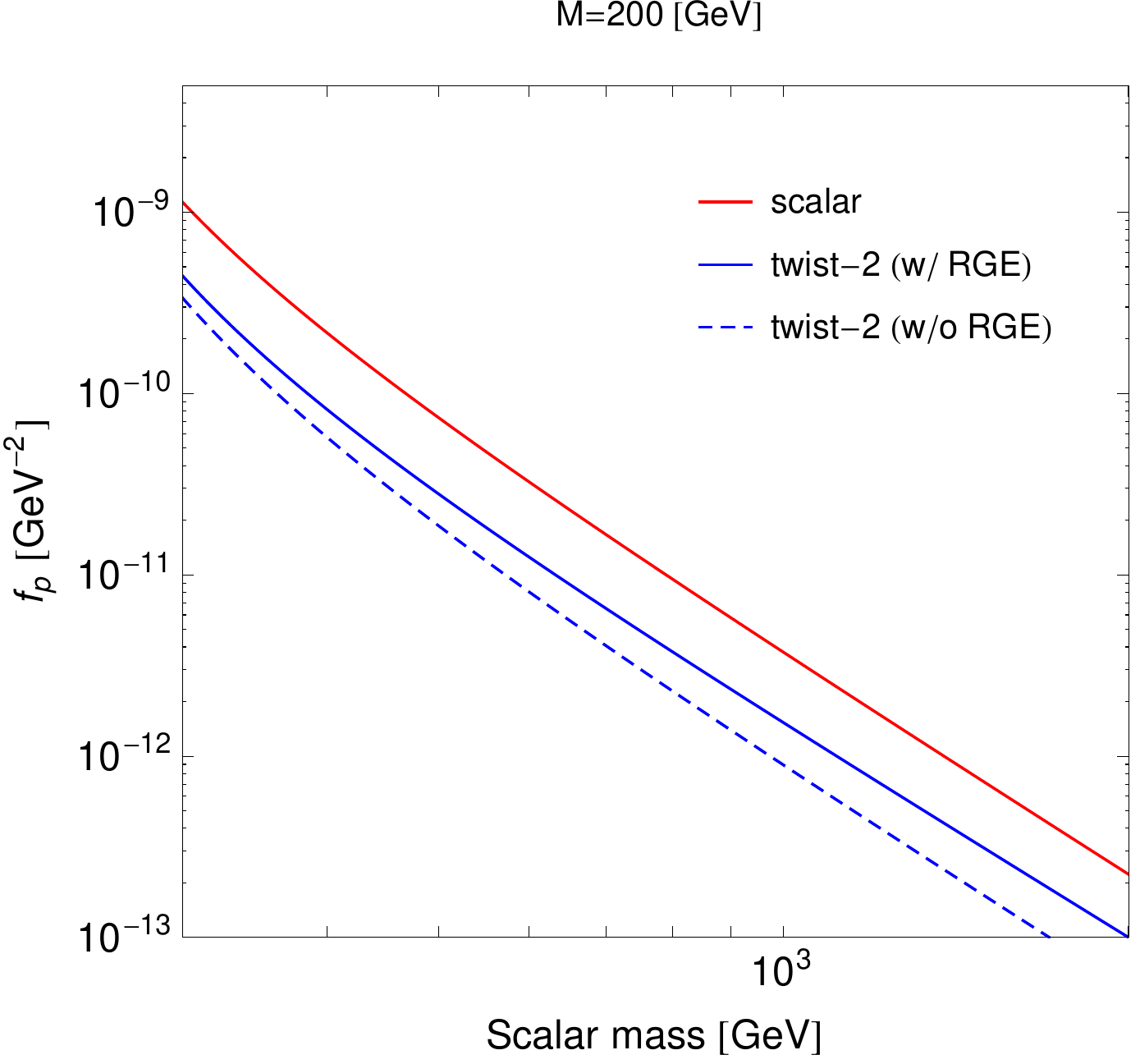}
 \label{fig:fp_3rd_rn}}
\hspace{0.05\textwidth}
\subfigure[Scattering cross section $\sigma_p$]
 {\includegraphics[clip, width = 0.45 \textwidth]{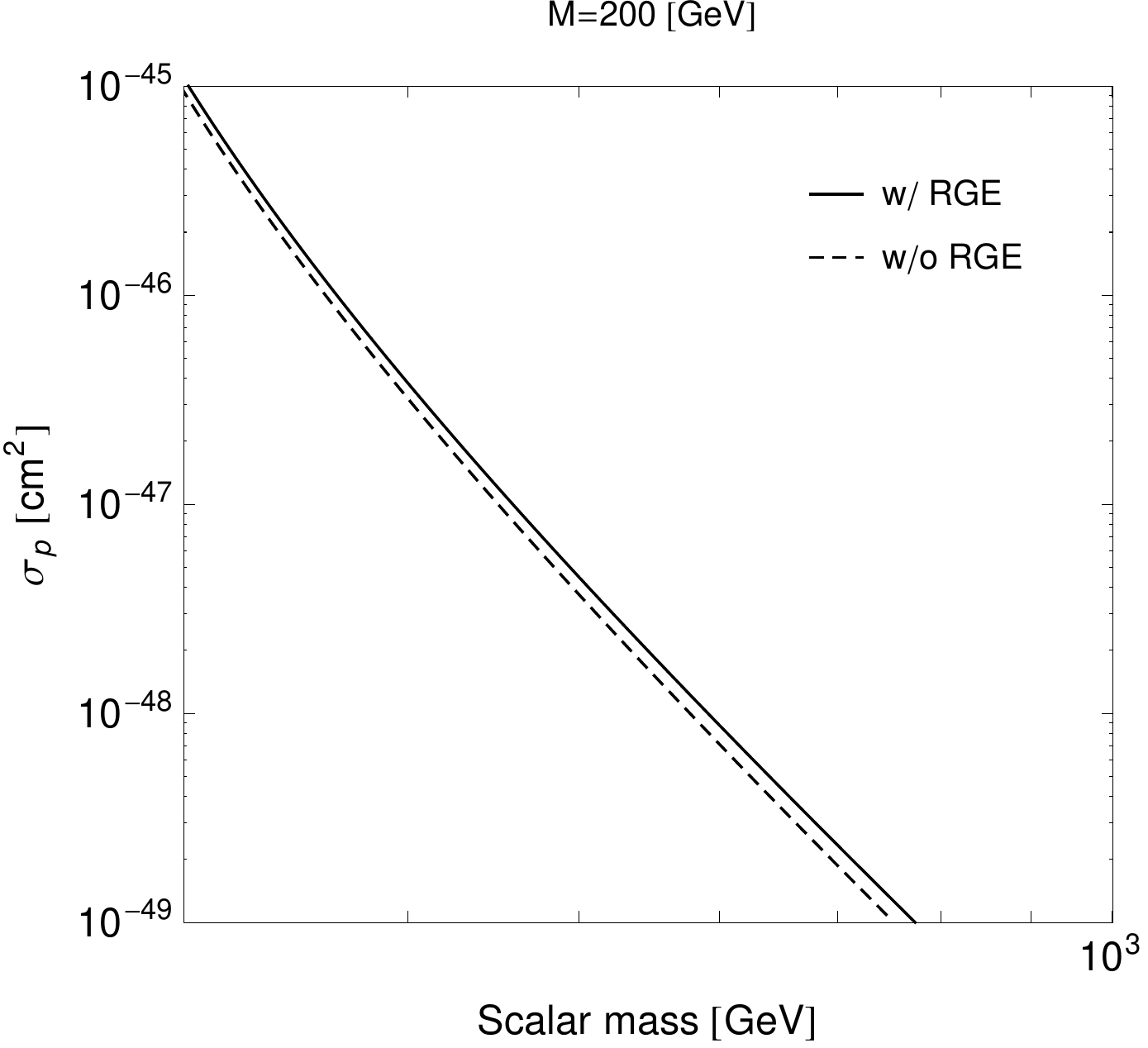}
 \label{fig:sicross_3rd_rn}}
\caption{(a) Each contribution to the WIMP-proton effective coupling
  $f_p$ as functions of the mediator mass $M_{\widetilde{q}}$. DM model
  adopted here is the same as Fig.~\ref{fig:twist2mz2gev}. Upper red
  (lower blue) line shows the contribution of the scalar-type
  (twist-2-type) operators. For the twist-2 contribution, solid and
  dashed lines show the results with and without the renormalization
  effects, respectively.  (b) WIMP-proton scattering cross section
  $\sigma_p$ as a function of $M_{\widetilde{q}}$. Solid and dashed
  lines show the results with and without the renormalization effects,
  respectively. In both plots, WIMP mass is set to be $M=200$~GeV.  }
\label{fig:results}
\end{center}
\end{figure}
%%%%%%%%%%%%%%%%%%%%%%%%%%%%%%%%%%%%%%%%%%%%%%%%%%%%%%%%%%

In Fig.~\ref{fig:fp_3rd_rn}, we show each contribution to the WIMP-proton
effective coupling $f_p$ as functions of the mediator mass
$M_{\widetilde{q}}$. We set the WIMP mass to be $M=200$~GeV. The upper red
(lower blue) line shows the contribution of the scalar-type
(twist-2-type) operators. For the twist-2 contribution, we show both 
calculations with and without the renormalization effects in solid
and dashed lines, respectively. By using the effective coupling, we then
evaluate the WIMP-proton scattering cross section $\sigma_p$. We plot it as a
function of $M_{\widetilde{q}}$ in Fig.~\ref{fig:sicross_3rd_rn}. Here
again the WIMP mass is set to be $M=200$~GeV. The solid and dashed lines
show the results with and without the renormalization effects,
respectively. We find that the renormalization effects
change the resultant value for the twist-2 contribution by more than
50\% when $M_{\widetilde{q}}\gtrsim 500$~GeV. In this case, the
scattering cross sections are modified by more than 20\%. The results
indicate that it is important to include the RGE effects, especially
when the colored mediators are much heavier than the electroweak scale.

%%%%%%%%%%%%%%%%%%%%%%%%%%%%%%%%%%%%%
\section{Conclusion and discussion}
\label{sec:conclusion}
%%%%%%%%%%%%%%%%%%%%%%%%%%%%%%%%%%%%%%

So far we have discussed a way of evaluating the WIMP-nucleon scattering
cross section at the leading order in $\alpha_s$ based on the effective
theoretical approach. We have considered a Majorana fermion, real scalar and vector bosons, and presented formulation for each case. Further, using a particular example with a Majorana fermion, we have shown that the renormalization effects may change the twist-2 contribution by more than 50\% when the colored mediators are much heavier than the electroweak
scale, which results in modification to the WIMP-nucleon scattering cross
section by ${\cal O}(10)$\%.

As shown in Fig.~\ref{fig:twist2mz2gev}, the calculation of the twist-2
contribution suffers from ${\cal O}(10)$\% uncertainty due to the
perturbation in $\alpha_s$. It is possible to reduce the uncertainty by
going beyond the leading-order calculation. In fact, we have already had
the higher-order results for the RGEs and the matching conditions at each
quark threshold, as commented in Sec.~\ref{sec:RGEs}. To complete the
next-to-leading order computation, however, we further need the
higher-order matching conditions between the full and effective theories
at the input scale. We defer the calculation as future work. In
addition, we expect that future lattice QCD simulations will much
improve the determination of the quark content in nucleon. These two
developments will enable us to evaluate the WIMP-nucleon scattering cross
sections with great accuracy.

Finally, we would like to emphasize that the prescription for the
computation of the WIMP-nucleon scattering cross section discussed in
this paper is quite systematic, and the formulation itself is almost
model-independent. The model-dependence is included in the Wilson
coefficients at the factorization scale, and the subsequent procedure is
similar in every case. Therefore, this method is suitable to be used in
general computational codes for the direct detection rate of DM, such
as {\tt micrOMEGAs} \cite{Belanger:2013oya} and {\tt DarkSUSY}
\cite{Gondolo:2004sc}.

%%%%%%%%%%%%%%%%%%%%%%%%%%%%%%%%%%%%
\section*{Acknowledgments}
%%%%%%%%%%%%%%%%%%%%%%%%%%%%%%%%%%%%

The work of J.H. is supported by Grant-in-Aid for Scientific research
from the Ministry of Education, Science, Sports, and Culture (MEXT),
Japan, No. 24340047 and No. 23104011, and also by World Premier
International Research Center Initiative (WPI Initiative), MEXT,
Japan. The work of R.N. and N.N. is supported by Research Fellowships
of the Japan Society for the Promotion of Science for Young
Scientists, No.~26$\cdot$3947 (R.N.) and No.~26$\cdot$8296 (N.N.).

%%%%%%%%%%%%%%%%%%%%%%%%%%%%%%%%%%%%%%%%%%%%%%%%%%%%%%%%%%%%%%%%%
\appendix
\section*{Appendix: gluon-loop contribution for real scalar boson DM}
%%%%%%%%%%%%%%%%%%%%%%%%%%%%%%%%%%%%%%%%%%%%%%%%%%%%%%%%%%%%%%%%%

In this Appendix, we give a result for the calculation of one-loop gluon
contribution in the case of real scalar boson DM, with quark masses kept
non-vanishing. Similar results have been already obtained for the cases
of Majorana fermion and real vector boson in Refs.~\cite{Hisano:2010ct}
and \cite{Hisano:2010yh}, respectively. 

%%%%%%%FIGURE%%%%%%%%%%%%%%%%%%%%%%

\begin{figure}[t]
\begin{center}
 \includegraphics[height=50mm,clip]{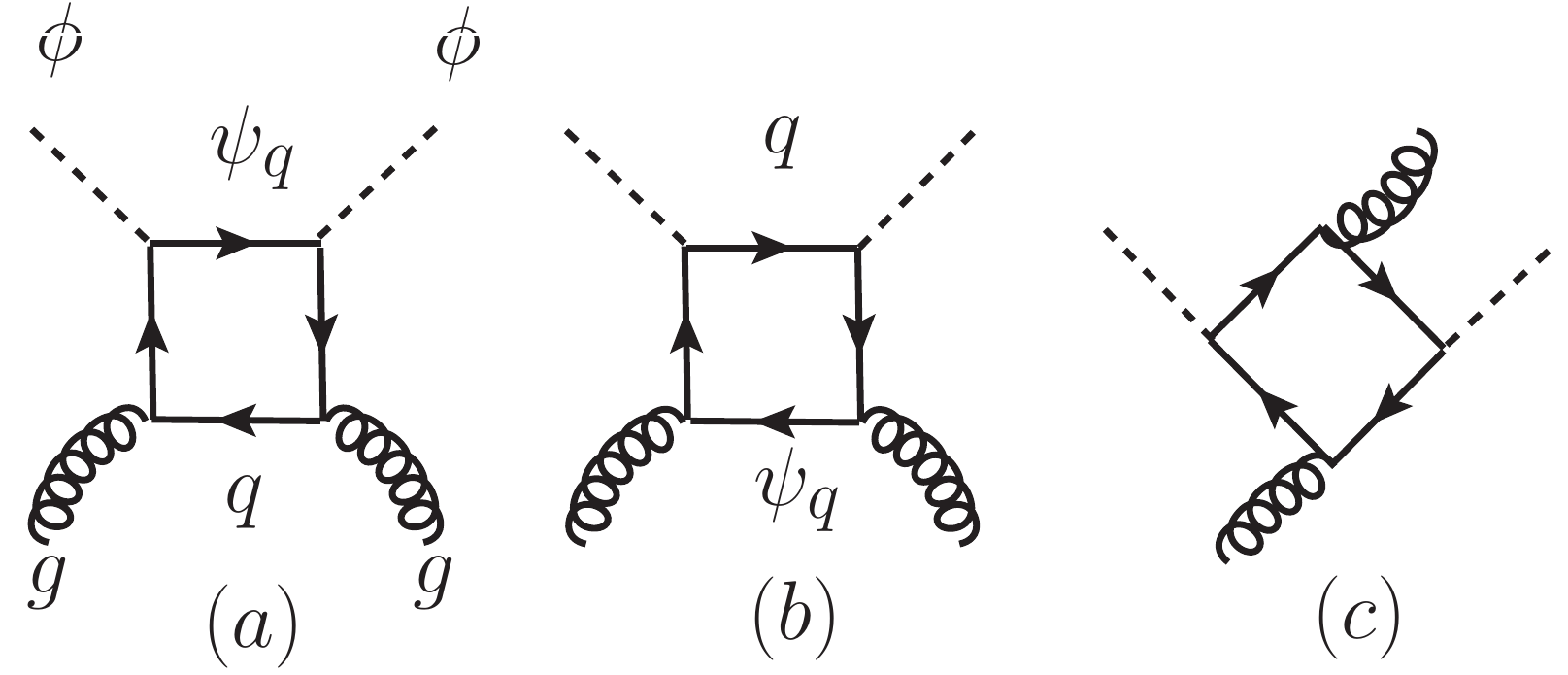}
\caption{One-loop contribution to the WIMP-gluon coupling.} 
\label{fig:scavacpol}
\end{center}
\end{figure}

%%%%%%%%%%%%%%%%%%%%%%%%%%%%%%%%%

We assume that the interactions of the real scalar boson with quarks and
the corresponding colored heavy fermions are described in terms of the
Lagrangian given in Eq.~\eqref{full:scalar}. 
The diagrams we consider here are illustrated in
Fig.~\ref{fig:scavacpol}. By evaluating the diagrams, we compute the
contribution of a heavy quark $Q$ to the coefficient of the gluon
scalar-type operator $C^g_S$ as 
\begin{align}
C^g_S|_Q&=\frac{1}{4}\sum_{i=a,b,c}
\left[(a^2_Q+b^2_Q)f^{(i)}_+(M;m_Q,m_{\psi_Q})
+(a^2_Q-b^2_Q)f^{(i)}_-(M;m_Q,m_{\psi_Q})  
 \right]~, 
\end{align}
where $f^{(i)}_{+}$ and $f^{(i)}_{-}$ $(i=a,b,c)$ correspond
to the contribution of the diagram $(i)$ in Fig.~\ref{fig:scavacpol}. 
They are given as follows:
\begin{align}
f^{(a)}_+(M;m_1,m_2)&\equiv -
 \frac{m^2_1m^4_2(M^2+m^2_1-m^2_2)}{\Delta^2}L \nonumber\\[3pt]
&-\frac{(-M^2+m_1^2+2m_2^2)\Delta + 6m_1^2m_2^2(M^2-m_1^2+m_2^2)}
{6\Delta^2}~,\\[3pt]
f^{(a)}_-(M;m_1,m_2)&\equiv 
\frac{m_1m_2^3\{\Delta +m_1^2(M^2-m_1^2+m_2^2)\}}{\Delta^2}L
\nonumber \\[3pt]
&-\frac{m_2\{
(-2M^2+m_1^2+2m_2^2)\Delta -6m_1^2m_2^2(M^2+m_1^2-m_2^2)
 \}}{6m_1\Delta^2} ~, \\[3pt]
f^{(b)}_+(M;m_1,m_2)&\equiv f^{(a)}_+(M;m_2,m_1) ~,\\[3pt]
f^{(b)}_-(M;m_1,m_2)&\equiv f^{(a)}_-(M;m_2,m_1)~,\\[3pt]
f^{(c)}_+(M;m_1,m_2)&\equiv
 \frac{-M^2+m^2_1+m^2_2}{2\Delta}-\frac{m^2_1m^2_2}{\Delta}L~,\\[3pt]
f^{(c)}_-(M;m_1,m_2)&\equiv
 \frac{2m_1m_2}{\Delta}-\frac{m_1m_2(-M^2+m^2_1+m^2_2)}{\Delta}L~,
\end{align}
with
\begin{align}
\Delta(M;m_1,m_2)&\equiv M^4-2M^2(m^2_1+m^2_2)+(m^2_2-m^2_1)^2~, \\[3pt]
L(M;m_1,m_2)&\equiv
\begin{cases}
\frac{1}{\sqrt{|\Delta|}}\ln\left( \frac{m^2_1+m^2_{2}-M^2+\sqrt{|\Delta
 |}}{{m^2_1+m^2_{2}-M^2-\sqrt{|\Delta |}}}\right) & (\Delta >0)  
 \\
 \frac{2}{\sqrt{|\Delta|}}\arctan\left( \frac{\sqrt{|\Delta
 |}}{{m^2_1+m^2_{2}-M^2}}\right) & (\Delta <0)
\end{cases}~~.
\end{align}
In particular, if $m_1\ll M,m_2$, the
above functions are approximated by
\begin{align}
 f_+^{(a)}&\simeq -\frac{2m_2^2-M^2}{6(m_2^2-M^2)^2} ~, \nonumber \\
 f_-^{(a)}&\simeq -\frac{m_2}{3m_1(m_2^2-M^2)} ~, \nonumber \\
 f_+^{(b)}&\simeq -\frac{1}{6(m_2^2-M^2)} ~, \nonumber \\
 f_-^{(b)}&\simeq 0~, \nonumber \\
 f_+^{(c)}&\simeq \frac{1}{2(m_2^2-M^2)}~, \nonumber \\
 f_-^{(c)}&\simeq 0~.
\end{align}
By using the expressions and the tree-level result in
Eq.~\eqref{CqSscalar}, one readily obtains the matching condition
\eqref{CgSscalar}.

%%%%%%%%%%%%%%%% References    %%%%%%%%%%%%%%%%%%%%
{}
%%%%%%%%%%%%%%%%%%%%%%%%%%%%%%%%%%%%%%%%%%%%%%%%%%%

\end{document}